\documentclass[prl,twocolumn,showpacs]{revtex4}

\usepackage{graphicx}%
\usepackage{dcolumn}% Align table columns on decimal point
\begin{document}

\title{Critical Behaviour of Thermal Relaxation in Disordered Systems}
\author{C. D. Mukherjee~}
\author{K. K. Bardhan}

\affiliation{Saha Institute of Nuclear Physics, 1/AF Bidhannagar, Calcutta
700 064, India}

\date{\today}

%\thanks[YY]{Corresponding author: {\tt e-mail: bardhan@cmp.saha.ernet.in}}

\begin{abstract}
At a composition far above the percolation threshold, the resistance of a
composite sample increases with time due to Joule heating as a constant
current of sufficiently large value is passed through the sample. If the
current is less than a certain breakdown current ($I_b$) the resistance
eventually reaches a steady value with a characteristic relaxation time
$\tau_h$. The latter diverges with current $I$ as $\tau_h \sim {(1-I^2/
{I_b}^2)}^{-z}$. The value of the exponent $z$ displays large fluctuations
leading to unusual scaling of the relaxation time. It is shown that the
results lead to important conclusions about the nature of breakdown
phenomena.
\end{abstract}


\maketitle

Breakdowns or fractures in random systems form an important class of
non-thermodynamic phase transitions\cite {bikas}. Much of the efforts\cite
{bikas,arcan,hansen,sornette,zapperi,andersen,pennetta,lamaignere,mbh}
to understand such irreversible phenomena naturally use the framework
already developed for thermodynamic critical phenomena although descriptions
remain far from complete. Consider, for example, the order of a breakdown
transition. Zapperi et al.\cite {zapperi} have suggested that the latter is
first-order as a function of the external field. The relevant 'order
parameter',  elastic constant or conductivity suffers a discontinuous change
from a finite value to zero at a breakdown point. Using a similar but
somewhat different model, Andersen et al.\cite {andersen} supported the idea
of a first order transition but only at small disorder. They predicted a
change to a second-order transition at higher disorder, thus indicating
presence of a tricritical point. On the experimental side, some recent
results on electrical failure in composites suggest rather conflicting
scenerios. When a sufficiently large current $I$ is passed through a
composite sample (a random binary mixture of a conductor and an insulator
\cite {stauffer}) with the conducting fraction $p$ typically far above the
percolation threshold $p_c \ll p$, Joule heating causes the sample
resistance to increase. If the current is more than the breakdown current
$I_b$, heating eventually leads to an irreversible electric breakdown. It
was found\cite {lamaignere} that the resistance of such a sample increases
as a power of time, $R \sim {(\tau_h(I)-t)}^{-0.65}$ when a constant current
$I>I_b$ is applied. The relaxation time (i.e. time-to-failure) $\tau_h$ also
exhibits a critical behaviour (see below). However, measurements\cite {mbh}
starting with small currents show that the sample conductance drops to zero
as soon as the current equals or exceeds the breakdown current, thereby
signalling a first-order transition.

In order to resolve this and other issues it would be useful to obtain
information about other relevant quantities near the transition. In this
paper, we report the results of our dynamic measurements of thermal
relaxation using currents \(less\) than, but upto, the breakdown current in
composite samples with \(varying\) amount of disorder. The observed
relaxation behaviour is reminiscent of the well-known phenomena of slowing
down\cite {hohenberg} near critical points in thermodynamic systems. The
relaxation times also exhibits an unusual scaling relation due to a strong
dependence on disorder. The present results together with the earlier
ones\cite {lamaignere} constitute, to our knowledge, the first full
description of a dynamic breakdown phenomenon \(both\) above and below the
critical parameter. We examine below various features including the
interplay of disorder and breakdown dynamics, which becomes significant
particularly in the regime of weak disorder $p_c \ll p \le 1$. Few related
experimental works include investigation of avalanche dynamics\cite {petri}
and strain-relaxation measurements in metal networks near the percolation
threshold\cite {ghosh}.

Measurements were performed in composites of carbon and wax (C-W)
under constant dc current condition at room temperature and the sample
resistance was monitored as a function of time during heating. The
preparation and characterisation of the system have been described
earlier\cite {cbb}. $p_c$ in C-W system is 0.76$\%$ (by volume). In the
tunneling regime\cite {kkb} close to $p_c$, the resistance \(decreases\)
with bias. It has been recently found out that there exists a Joule regime
in, or a conducting fraction $p_J$ above, which the resistance always
\(increases\) with bias\cite {mbh}. The nominal carbon fractions of the
samples ranging from $4.5$ to $10\%$ were above or near $p_J$ which is
approximately 4.5$\%$\cite {nmb3}. Samples prepared with larger carbon
fractions were mechanically unstable. On the other hand, samples with
lower carbon fractions had tunneling effect offsetting the effect of Joule
heating (see Fig. 1 of Ref. \onlinecite {mbh}). The geometry of current flow
in a cylinderical sample is illustrated in the inset in Fig. 1. Various
properties of the samples used in the present work are given in Table I. In
a typical measurement, the current corresponding to a certain value $I$
would be turned on at time $t=0$ and the bias across the sample be measured
at an interval of time (usually 250 msec) until it reached a steady value
when a balance between dissipation and generation of heat within the sample
is established. The sample was then allowed to cool for up to 2 hours before
measurements would be repeated with an another current, not exceeding the
breakdown current. The relaxation during cooling from a hot steady state was
also measured. It exhibited an anamalous behaviour. This will be reported
elsewhere.

\begingroup
\squeezetable
\begin{table}
\caption{Various properties of the samples used. $p$ is the carbon fraction
by volume ($\%$). $R_o$, $\rho_o$ and $\tau_{ho}$ are the resistance,
resistivity and relaxation time at zero current. $I_b$ is the breakdown
current and $z$ is the exponent in Eq. (\ref {eq:tau}).}
%\label{tab:example}
\begin{ruledtabular}
\begin{tabular}{ccccccdd}
Sample &\(p\)\footnote{Nominal value.} &Height &$R_o$ &$\rho_o$ &$\tau_{ho}$
&\mbox{\(I_b\)} &\mbox{\(z\)} \\
  No.  &                               &(mm)   &\mbox{\((\Omega )\)}
&\mbox{(\(\Omega\)cm)} & (sec) &\mbox{(mA)} &        \\
\hline
    1 &7.5 &6.0     &422  &567  &495  &31.5   &0.175        \\
    2 &4.5 &4.1     &287  &561  &145  &19.2   &0.187        \\
    3 &7.5 &4.1     &139  &280  &168  &73     &0.30        \\
    4 &4.5 &2.7     &~77  &214  &455  &83.1   &0.076       \\
    5 &7.5 &2.8     &~73  &191  &113  &112    &0.18        \\
    6 &7.5 &4.2     &~94  &188  &156  &95     &0.15       \\
    7 &7.5 &10x4x2\footnotemark[2] &184  &15.2  &~67 &18.7 &0.23 \\
    8 &10  &8.3     &5.56 &5.26 &194  &135    &0.173 \\
\end{tabular}
\end{ruledtabular}
\footnotetext[2]{This sample is ribbon-shaped but all other samples are
cylinders of 10mm diameter.}
\end{table}
\endgroup

Figure 1 shows a typical relaxation curve (a) during heating, obtained from
a cylinderical sample ($\#$3). The curve is well described by a simple
exponential process indicated by the solid line in Fig. 1. It may be
recalled\cite {campbell} that the relaxation function $f(t)$ is almost
universally given by $f(t) \sim \exp [{-(t/\tau)}^{\alpha}]$. Here $t$ is
time, $\tau$ is a relaxation time constant and $\alpha$ is an exponent. The
exponential relaxation corresponding to $\alpha = 1$ occurs mostly in simple
systems such as homogeneous ordered solids\cite {rem1}. A simple exponential
in the present case may be attributed to a long-time relaxation behaviour.
In systems with characteristic time scales such as percolating networks, the
relaxation function may change from a stretched exponential at short times
to a simple exponential at long times \cite {havlin,ghosh}. For samples with
large $p \gg p_c$ such as  the ones used here, the cross-over times should
be quite small. Close to $p_c$, $\alpha$ was determined to be about 0.8\cite
{ghosh} in elastic relaxation in dilute metal foils, and about 0.4\cite
{nmb97} in voltage relaxation in the same system as the present one.

\begin{figure}
\includegraphics{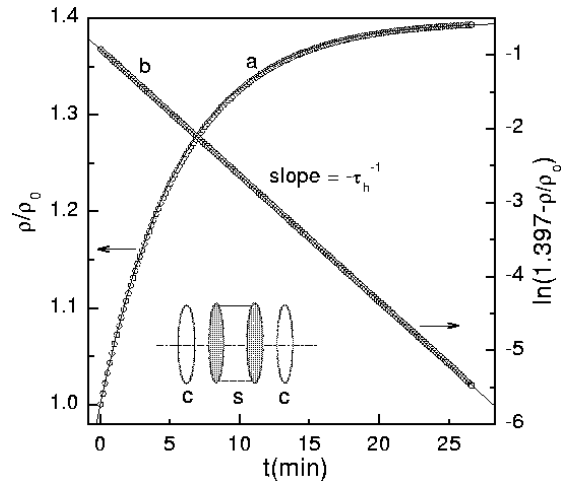}
\caption {A typical resistance relaxation curve (a) of a cylinderical sample
($\#$3) heated by passing a constant current of 68.5mA. Many data points
have been omitted for clarity. The solid line is a fit to a simple
exponential $\rho /\rho_o = 1.397- 0.397 \exp (-t/\tau_h)$. The same data
are also shown in log-linear plot (b). The slope of the straight line yields
negetive inverse of the relaxation time $\tau_h$=348s. The inset is an
assembly diagram showing copper foils (c) used as electrodes and attached
to the sample (s) with silver paints.} \label{fig:1}
\end{figure} %

\begin{figure}
\includegraphics{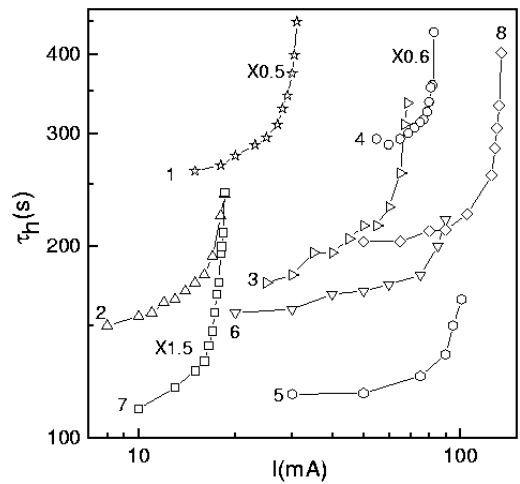}
\caption {Relaxation time vs. current in various samples as in Table I. Some
data have been translated vertically by factors as marked.} \label{fig:2}
\end{figure} %

The relaxation time in several samples of different compositions and
dimensions (see Table I) against current are shown in Fig. 2. $\tau_h$ of a
sample was found to diverge as a power-law
\begin{equation}
\tau_h = \tau_{ho} {\mid \epsilon \mid}^{-z}       \label{eq:tau}
\end{equation}
Here, $\tau_{ho}$ is a prefactor, $\epsilon = (I^2/{I_b}^2-1)< 0$ and $z$ is
a dynamical exponent. $\tau_{ho}$, $I_b$ and $z$ were treated as fitting
parameters and its values are listed in Table I. $\tau_{ho}$ is really the
relaxation time at zero current. It depends on sample dimensions (see Eq.
\ref {eq:tauhomo}) but more importantly, monotonically increases with
disorder. In fact, $\tau_{ho}$ diverges as $p$ approaches $p_c$. Such
disorder-induced divergence of relaxation time has been qualitatively
observed by Ghosh et al.\cite {ghosh}. The breakdown current $I_b$ scales
with $R_o$ as $I_b \sim {R_o}^{-0.44}$\cite {mbh}. $R_o$ is the resistance
at zero current.

The exponent $z$ thus determined was not a constant even within experimental
unceratinities of 5$\%$. It exhibits an unusual sample-to-sample fluctuation
as seen in Fig. 3 where it is plotted against a quantity $r$ as a measure of
disorder. (It is not practical to use $p$ due to uncertainities in its
values). $r$ is defined as the ratio of the zero-current sample resistivity
and the resistivity of the pure conducting phase ($\sim 10^{-2} \Omega \rm
{cm}$). $r$ is 1 at zero disorder ($p=1$) and increases monotonically with
disorder as $p$ decreases from 1. Within the range of disorder (two orders
of magnitude) investigated in this work, the exponent has an average value
of $0.184 \pm 0.06$ i.e 33$\%$ uncertainities. The highest value (0.3) is
about four times the lowest one (0.076). Without these two extreme values,
uncertainities reduce to 17$\%$ which is still beyond the experimental
error. Clearly, the relaxation exponent $z$ for $I<I_b$ is
disorder-dependent. In fact, one can classify the breakdown-related
exponents, both theoretical and experimental, into two groups: 1)
disorder-dependent and 2) disorder-independent or 'universal'. Examples of
the first group include, besides $z$ of this work, the ones associated with
broken bonds in lattice models\cite {batrouni} while those in the second
group include the size exponents in Table I of Ref. \onlinecite {hansen},
roughness exponent\cite {batrouni}, relaxation exponent $z$ for $I>I_b$ (see
below). A disorder-dependent exponent is usually derived from measurements
involving a $single$ sample and hence, fluctuations in its values possibly
reflect lack of self-averaging property.

\begin{figure}
\includegraphics{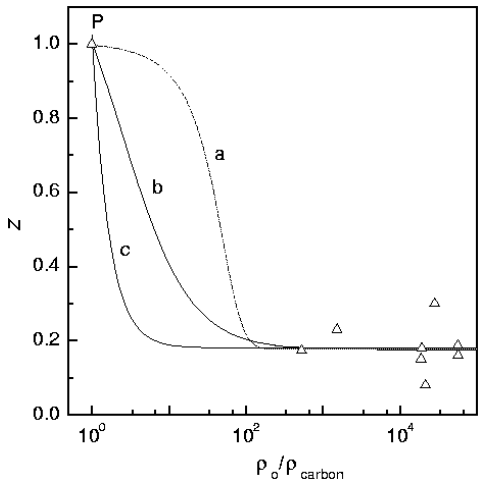}
\caption {The exponent $z$ vs. the disorder parameter $r=\rho_o/{\rho}_
{carbon}$ for $I<I_b$. The point P(1,1) is a theoretical one corresponding
to the pure conductor. See text for furthur discussion.} \label{fig:3}
\end{figure}

To show how well Eq. \ref {eq:tau} represents the relaxation time data, one
notes that plotting ${(\tau_h/\tau_{ho})}^ {1/z}$ against $I^2/{I_b}^2$
should lead to data collapse. This is indeed seen in Fig. 4(open symbols,
$I^2/{I_b}^2<1$). The solid line is a fit to (\ref {eq:tau}), which is
excellent even at low currents. Thus, Eq. \ref {eq:tau} holds not only near
$I_b$ but also for the entire range of current $I<I_b$. Goodness of the data
collapse near $I_b$ is highlighted in the plot in the inset. Staight lines
have slopes of unity. To complete the description of the dynamic breakdown,
time-to-failure data (closed symbols) obtained above the breakdown point
($\epsilon>0$) by Lamaignere et al.\cite {lamaignere} are also shown in the
figure. In this case, $\tau_{ho}$ is fixed arbitrarily to display the data
within the graph and can not be interpreted as the relaxation time at zero
current as in the case $\epsilon <0$. The dashed line is again a fit to
(\ref {eq:tau}) with $z=1$. The single function fits reasonably well all the
data whereas previously, different functions were used to fit different
ranges of data. The function $\tau_h \sim (I/I_b-1)^{-2}$ as considered in
Ref. \onlinecite {lamaignere} near $I_b$ is also shown in Fig. 4(dotted
line). As seen, it does not fit the whole set of data. It is important to
note that the value of the exponent $z$ (i.e. 1) for $\epsilon > 0$ is
$independent$ of disorder in contrast to the varying exponent for $\epsilon
< 0$. This is ensured by Eq. \ref {eq:tau} and the requirement that $\tau_h$
must vary as $I^{-2}$ at large currents \cite {lamaignere} as dissipation of
heat becomes negligible at large currents.

\begin{figure}
\includegraphics{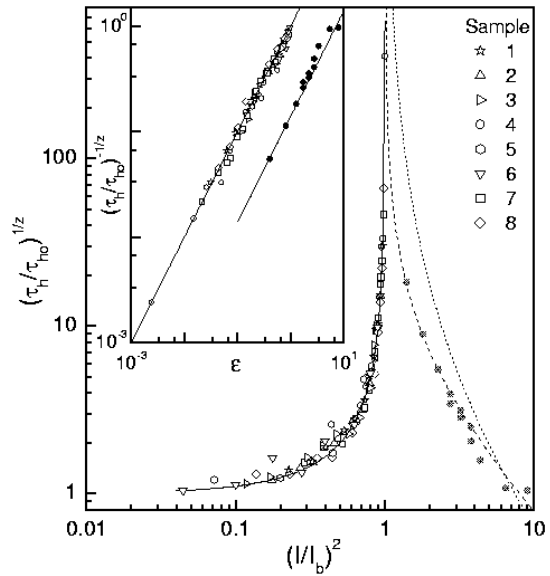}
\caption {Scaling behaviour of relaxation time below and above the
breakdown point at \(I=I_b\). The solid and dashed lines are fits to
${(\tau_h / \tau_{ho})}^ {1/z} = {(\mp \epsilon)}^{-1}$ respectively. The
data corresponding to solid symbols are from Ref. \onlinecite {lamaignere}.}
\label{fig:4}
\end{figure} %

It is now useful to summarize characteristics below and above the breakdown
point $I=I_b$. (i) For \(\epsilon<0\), the resistance ratio $R/R_o$ (see,
for example, Fig. 1 of Ref. \onlinecite {mbh}) smoothly increases with
current to a limiting value (or breakdown ratio) \(\Upsilon \approx
1.5\)\cite {nmb3} at \(\epsilon=0\). $R= R(t \rightarrow \infty)$ is the
steady resistance at a given current. It is best fitted by an expression
$R/R_o= \Upsilon + a_1 \epsilon + a_2 {\epsilon}^2 + a_3 {\epsilon}^3$ that
has a finite slope $a_1$ at \(\epsilon=0\). For \(\epsilon>0\), there is no
steady state. On the other hand, $\tau_h$ diverges as a power-law of
\(\epsilon\) on both sides; ii) $z$ is not same on both sides as discussed
above. This is in contrast to equality of the exponents on both sides, that
is expected from scaling hypotheses in both static and dynamic critical
phenomena in thermodynamic systems; iii) While the scaling in Fig. 4 is
itself an expression of 'universality' \cite {hansen}, the fluctuations in
$z$ for \(\epsilon<0\) renders the notion of an universality class
unteneble; iv) At \(\epsilon<0\), both $R$ and $\tau_h$ are \(reversible\)
with respect to current. This rules out any local breakdown $before$ the
global breakdown occurs. For, any irreversible change in the microstructure
should also lead to irreversibility in those quantities. This is
incompatible with the picture that emerges in breakdown models with
$annealed$ disorder\cite {arcan,sornette,andersen,zapperi, pennetta} where
the global breakdown is preceded by increasing bursts of irreversible bond
breaking. The composite samples can be considered belonging to a system with
quenched disorder for \(\epsilon<0\), and a system with annealed disorder
for \(\epsilon>0\). In view of all these, it is difficult to avoid the
conclusion that the usual thermodynamic classification of phase transition
is inadequate to describe breakdown phenomena.

The presence of disorder is expected to have a profound effect on the
breakdown processes. One of its manifestation lies in the conceptual
difficulties of taking the limit of disorder going to zero\cite {andersen}.
This is aptly illustrated in the present case. Let us first consider
divergence of the relaxation time in a homogeneous medium corresponding to
zero disorder. In fact, it has been long discussed\cite {jaeger} albeit
couched in the language of stability. The Joule heating under a constant
current $I$ in a medium with a positive temperature coefficient of
resistance $\beta$ leads to either a steady state or breakdown depending on
whether the heat generated is removed quickly enough or not. Full solutions
of heat flow have been obtained in some regular geometries\cite {jaeger}
where $\tau_h$ is given by
\begin{equation}
\tau_h \sim l^2 {(\kappa a - bI^2)}^{-1}  \sim l^2\kappa {(1-I^2/{I_b}^2)}
^{-1} \label{eq:tauhomo}
\end{equation}
where ${I_b}^2= \kappa a/b$, $\kappa$ is thermal diffusivity, and $l$ is
the smallest distance for flow of heat. $a,b$ are constants that depends on
the boundary conditions and various material constants including $\beta$.
Clearly, $z = 1$ in the case of zero disorder ($p=1$). In passing let us
note that according to Eq. \ref {eq:tauhomo}, $I^2$ rather than $I$ is the
proper variable to use in problems involving the Joule heating.
Consideration of temperature-coupled resistance is an essential ingradient
in obtaing (\ref {eq:tauhomo}). Its absence in the dynamic fuse model of
Sornette et al.\cite {sornette} gives rise to a relaxation time
\(independent\) of current.

As the disorder is reduced ($p \rightarrow 1$), the relaxation exponent
increases from $\sim 0.18$ to 1. Thus, the dependence of $z$ on disorder is
much stronger than the fluctuations in its values suggest. Presently, there
is no theory of $z$ for $p<1$. An interesting question arises at this point
as to how $z$ from a lower value at higher disorder would approach 1 at the
point P in Fig. 3. Three possible curves are shown schematically. The curve
$a$ is intuitive and has a zero slope at $P$. This means that the system
having a small disorder can be simply considered as a homogeneous one with
an effective thermal diffusivity. This may be possible if the melting point
of the conductor is \(less\) than that of the insulating matrix. The curve
$c$ corresponds to the situation where even a slight disorder leads to an
abrupt fall in $z$. This may happen if the melting point of the conductor is
\(greater\) than that of the insulating matrix. The curve $b$ has a finite
slope at P. Clearly, furthur theoretical efforts are necessary for
understanding of $z$ as a function of disorder. Similar problem\cite {mbh}
concerning the breakdown ratio \(\Upsilon\) exists near $p \sim 1$.

In conclusion, a detailed description of the dynamic electric breakdown
driven by external current, both below and above the breakdown current, was
presented. It was shown that the classification of thermodynamic phase
transitions is inadequate for breakdown transitions. The present models
having annealed disorder lack reversibility found in composites.

\begin{acknowledgments}
We acknowledge several discussions with Bikas Chakrabarti and the assistance
of Arindam Chakrabarti in the preparation of C-W samples and processing
data.
\end{acknowledgments}


\begin{thebibliography}{22}
\bibitem[{rem()}]{rem1}
\bibinfo{note}{When metal blocks were used as electrodes, relaxation data
  required at least two exponentials with different relaxation times for
  fitting.}
\end{thebibliography}
\end{document}